\documentclass[12pt]{article}
\usepackage{graphicx}
\DeclareGraphicsExtensions{.pdf}
\usepackage{float} %Include figure filesusepackage{graphicx} %Include figure files
\usepackage{amsmath}
\usepackage{amsfonts}   
\usepackage{amssymb}

\begin{document}
\date{}
%%%%%%%%%%%%%%%%%%%%
\title{{\bf{\Large Quantum fluctuations and thermal dissipation in higher derivative gravity}}}
%%%%%%%%%%%%%%%%%%%%
\author{
 {\bf {\normalsize Dibakar Roychowdhury}$
$\thanks{E-mail:  dibakarphys@gmail.com, dibakar@cts.iisc.ernet.in}}\\
 {\normalsize Centre for High Energy Physics, Indian Institute of Science, }
\\{\normalsize C.V. Raman Avenue, Bangalore 560012, Karnataka, India}
%\\[0.3cm]
}
%\date{}

\maketitle
%%%%%%%%%%%%%%%%%%%%%%%%%%%%%%%%%%%%%%%%%%%%%%%%%%%%%%%%%%%%%%%%%%%%%%%%%%%%
\begin{abstract}
In this paper, based on the $ AdS_{2}/CFT_{1} $ prescription, we explore the low frequency behavior of quantum two point functions for a special class of strongly coupled CFTs in one dimension whose dual gravitational counterpart consists of \textit{extremal} black hole solutions in higher derivative theories of gravity defined over an asymptotically AdS space time. The quantum critical points thus described are supposed to correspond to a very large value of the dynamic exponent ($ z\rightarrow \infty $). In our analysis, we find that quantum fluctuations  are enhanced due to the higher derivative corrections in the bulk which in turn increases the possibility of quantum phase transition near the critical point. On the field theory side, such higher derivative effects would stand for the corrections appearing due to the finite coupling in the gauge theory. Finally, we compute the coefficient of thermal diffusion at finite coupling corresponding to Gauss Bonnet corrected charged Lifshitz black holes in the bulk. We observe an important crossover corresponding to $ z=5 $ fixed point. 
\end{abstract}

%%%%%%%%%%%%%%%%%%%%%%%%%%%%%%%%%%%%%%%%%%%%%%%%%%%%%%%%%%%%%%%%%%%
\section{Overview and Motivation}
%%%%%%%%%%%%%%%%%%%%%%%%%%%%%%%%%%%%%%%%%%%%%%%%
Dissipative dynamics of a massive charged quark moving through a strongly coupled hot $ \mathcal{N}=4 $ SYM plasma has been one of the major focus of theoretical investigations during the past one decade \cite{Herzog:2006gh}-\cite{Gubser:2006bz}. Under a holographic set up, an infinitely massive quark could be thought of as an end point of an open string hanging from the asymptotic AdS boundary to the horizon of the black hole. On the other hand, a quark with finite mass and in the fundamental representation of the gauge group could be thought of as the end point of an open string stretched in between a flavor brane (that fills the space time at a finite radial distance) and the horizon\footnote{At this stage it is noteworthy to mention that through out our analysis we would assume that the open string does not back react on the background geometry.}. 

The dissipative dynamics, in particular the effect of drag force on the massive charged quark moving through the hot viscous plasma has been extensively investigated in \cite{Herzog:2006gh}-\cite{Gubser:2006bz} and subsequently in several other works \cite{Chernicoff:2006hi}-\cite{Giataganas:2013zaa}. The outcome of these analysis is that due to the viscous drag, the quark constantly loses its energy as it passes through the hot plasma. The corresponding momentum flux flows down towards the horizon along the world sheet. As a result of this, in order to keep the quark moving with a constant velocity $ v $, one needs to feed energy to the quark at a constant rate and which is achieved by applying a constant electric field at the boundary \cite{Herzog:2006gh}. For some excellent reviews on heavy ion collisions in AdS/CFT see \cite{CasalderreySolana:2011us}-\cite{Kovchegov:2011ub}.

Recently, the dissipative dynamics associated with the quantum critical systems has been extensively investigated in the context of nonrelativistic field theories with arbitrary dynamic exponent ($ z $) \cite{Tong:2012nf}-\cite{Edalati:2012tc}. In \cite{Tong:2012nf}, authors study the dissipative quantum dynamics associated with quantum critical systems at a Lifshitz fixed point.  These analysis has been further extended for quantum critical systems with hyperscaling violation ($ \theta >0 $) in \cite{Edalati:2012tc}. In their analysis \cite{Tong:2012nf} of the zero temperature dissipative forces associated with the quantum critical point in the presence of a massive probe particle, the authors had found out a crossover corresponding to the $ z=2 $ fixed point which they identified as a consequence of the fact that the inertial mass ($ m $) of the particle becomes irrelevant for $ z>2 $ which therefore does not contribute to the low energy dynamics of the quantum two point function corresponding in this particular regime of $ z $. In theories with hyperscaling violation, such a crossover was observed for $ z+\frac{2\theta}{d} \geq 2$ \cite{Edalati:2012tc}, where $ \theta $ is the so called hyperscaling violating exponent and $ d $ is the number of spatial directions for the spacetime.

One of the fascinating outcomes of these analysis is that in the limit $ z \rightarrow \infty $, the  low frequency (or energy) behavior of the quantum two point function thus computed for the dual field theory turns to be extremely simple and quite illuminating namely \cite{Edalati:2012tc},
\begin{eqnarray}
<X(\mathfrak{w})X(0)> ~\sim \frac{1}{\mathfrak{w}}\label{Q}
\end{eqnarray}
regardless of the value of the hyperscaling violating exponent ($ \theta $) and the dimensionality ($ d $) of the spacetime. In a holographic construction, the above result (\ref{Q}) could be realized by considering extremal RN black holes in the bulk where the dual theory in its deep IR flows towards a one dimensional CFT  which essentially describes a quantum critical point with $ z=\infty $ \cite{Edalati:2012tc}. In other words, the above result (\ref{Q}) corresponds to the low frequency behavior of quantum correlators corresponding to a quantum critical system at large values of $ z $ where the dual counterpart consists of extremal AdS black holes in the bulk whose near horizon geometry turns out to be $ AdS_{2}\times \mathbb{R}^{d-2} $. Such an observation is not quite surprising due to the fact that the spacetime with Lifshitz scaling goes over to $ AdS_{2}\times \mathbb{R}^{d-2} $ in the large $ z $ limit \cite{Edalati:2012tc}. This could be checked momentarily as follows. Consider the following Lifshitz metric in $ d $ dimensions \cite{Tong:2012nf},
\begin{eqnarray}
ds^{2}=- r^{2z}dt^{2}+\frac{dr^{2}}{r^{2}}+r^{2} d\textbf{x}^{2}.\label{L1}
\end{eqnarray}
Now consider the following coordinate transformation namely,
\begin{eqnarray}
\varrho =\frac{1}{r^{z}}.\label{L2}
\end{eqnarray}
Substituting (\ref{L2}) into (\ref{L1}), we finally obtain,
\begin{eqnarray}
ds^{2}=\frac{1}{z^{2}\varrho^{2}}(-d\tilde{t}^{2}+d\varrho^{2})+\frac{1}{\varrho^{2/z}}d\textbf{x}^{2}
\end{eqnarray}
which precisely goes over to $ AdS_{2}\times \mathbb{R}^{d-2} $ in the large $ z $ limit. The bottom line is therefore any analysis performed over (\ref{L1}) should be equivalent to an analysis performed over $ AdS_{2}\times \mathbb{R}^{d-2} $ in the large $ z $ limit.  These analysis therefore suggest that the low energy behavior of the correlation function should essentially be captured by the near horizon structure of the extremal black hole in the bulk. The basic characteristic feature of the above result (\ref{Q}) thus turns out to be universal as long as the extremal black holes are concerned. This observation is therefore quite similar in spirit to that of the shear viscosity calculations in the context of extremal black holes \cite{Edalati:2009bi}-\cite{Paulos:2009yk}. 

The purpose of the present article is precisely to push the above idea in the context of higher derivative theories of gravity and to see whether there are any effects of higher derivative corrections on the quantum two point function evaluated at the end point of the string corresponding to $ z=\infty $ fixed point. In the language of Gauge/gravity duality, the higher derivative corrections in the bulk correspond to finite $ \lambda_{t} $  corrections on the gauge theory side, where  $ \lambda_{t}(=g_{YM}^{2}N) $ is the so called t' Hooft coupling corresponding to $ \mathcal{N}=4 $ SYM plasma. In our analysis, following the original approach of \cite{Edalati:2012tc}, we compute quantum two point function corresponding to $ z=\infty $ fixed point considering extremal black holes in higher derivative theories of gravity whose near horizon structure essentially turns out to be $ AdS_{2}\times \mathbb{R}^{d-2} $. Following the previous discussion, it is therefore quite natural to expect that the low frequency fluctuations at the end point of the string must exhibit some unique feature as mentioned in (\ref{Q}). In addition to that one might expect some finite higher derivative corrections to quantum correlators which should be smoothly mapped to (\ref{Q}) in the limit of the vanishing coupling. The source of these higher derivative corrections could be easily anticipated from the induced metric on the world sheet geometry which thereby affects the dynamics of the scalar fields (fluctuations) propagating over the string world sheet.

Looking at the present analysis from a more general perspective, one could think of it as quite similar in spirit to that of the earlier hydrodynamic analysis performed in the context of extremal black holes\cite{Edalati:2009bi}-\cite{Paulos:2009yk}. Since the low frequency behavior of the hydrodynamic fluctuations are essentially captured by the near horizon geometry of the black holes in the bulk, as a result the $ \eta/s $ ratio turns out to be universal both in the Einstein as well as in the Gauss Bonnet (GB) gravity \cite{Edalati:2009bi}-\cite{Paulos:2009yk}. The present analysis is quite relevant in this sense as the low frequency behavior of the quantum fluctuations are also captured by the near horizon data in the bulk. Surprisingly, unlike the hydrodynamic scenario, we notice some finite higher derivative corrections to the quantum correlation function corresponding to $ z=\infty $ fixed point which turns out to be additive. The significance of this result is therefore the following: The higher derivative corrections on the gravity sector enhance the quantum fluctuations near the quantum critical point which thereby increases the possibility of the corresponding quantum phase transition.

The organization of of the present paper is the following:  In Section 2, we compute two point quantum correlation function corresponding to $ z=\infty $ fixed point both in the Gauss Bonnet (GB) \cite{Cai:2009zn} as well as in the third order Lovelock gravity \cite{Ge:2009ac}. It is noteworthy to mention that in the limit of the vanishing coupling, our result smoothly matches to that with the earlier findings in \cite{Edalati:2012tc}. In Section 3, for the sake of completeness, we compute the thermal diffusion coefficient corresponding to charged Lifshitz black holes in GB gravity \cite{Pang:2009pd}. We observe an important crossover at certain values of the dynamic exponent ($ z $). Finally, we conclude in Section 4.

%%%%%%%%%%%%%%%%%%%%%%%%%%%%%%
\section{Fluctuations: $ AdS_2/CFT_1 $ correspondence}
%%%%%%%%%%%%%%%%%%%%%%%%%%%%%%%%%%%%%%%%%%%%%%%%%%%%%%%%%%%%%%%%
\subsection{Gauss-Bonnet gravity}
In this section, our aim is to compute the two point quantum correlation function associated with the quantum critical system\footnote{Note that the UV fixed point of the boundary field theory is described in terms of a $ CFT_4 $ which is dual to $ AdS_5 $. On the other hand, the in the deep IR the theory flows to a $ CFT_{1} $ which is dual to $ AdS_2 $.} (corresponding to the $ z=\infty $ fixed point) in the presence of a massive charged (quark like) particle at the boundary. What we would essentially study is the linear response of the system when perturbed due to some external (electrical) force. In order to compute the two point correlation function, we would assume the so called fluctuation dissipation theorem \cite{deBoer:2008gu} which relates the two point correlation with the imaginary part of the response function. In our analysis we are basically interested to compute the two point quantum correlation function at finite coupling. In other words, we look for finite  $ \lambda_{t}$ corrections to the quantum correlation function. In order to explore such non trivial effects at the boundary, in the dual gravitational description one needs to go beyond the usual notion of two derivative Einstein gravity and include higher derivative corrections to the usual Einstein-Hilbert action. 

Keeping all these facts in mind, in our analysis we start  with a minimal introduction to the dual gravity set up in the bulk which is essentially described in terms of extremal Gauss-Bonnet(GB) RN black holes in ($ 4+1 $) dimensions. The corresponding action turns out to be \cite{Cai:2009zn},
\begin{eqnarray}
S&=&\frac{1}{2\kappa^{2}}\int d^{5}x \sqrt{-g}\left[R+\frac{12}{l^{2}}+\frac{\lambda l^{2}}{2}\mathcal{L}_{GB}-\frac{\kappa^{2}}{2} F_{\mu\nu}^{2} \right]\nonumber\\
\mathcal{L}_{GB}&=& R_{abcd}R^{abcd}-4R_{ab}R^{ab}+R^{2}\label{E1}
\end{eqnarray}
where $ \lambda $ is the GB coupling and $ \kappa^{2}=8 \pi G $.
The extremal background that results from the above action (\ref{E1}) could be expressed as \cite{Cai:2009zn},
\begin{eqnarray}
ds^{2}&=& \frac{u^{2}r_H^{2}}{l^{2}}\left(-f(u)\mathcal{N}^{2} dt^{2}+dx_{i}^{2} \right)+\frac{l^{2}du^{2}}{u^{2}f(u)}\nonumber\\
f(u)&=& \frac{1}{2\lambda}\left[1-\sqrt{1-4\lambda\left(1-u^{-2} \right)^{2}(1+2u^{-2}) } \right] \nonumber\\
\mathcal{N}^{2}&=&\frac{1}{2}(1+\sqrt{1-4\lambda}),~~A=\mu (1-u^{-2})dt,~~\mu=\frac{\sqrt{3}r_H \mathcal{N}}{\kappa l}\label{E2}
\end{eqnarray}
where $ u (=r/r_H)$ is a dimensionless radial variable such that the horizon is placed at $ u=1 $ and the boundary is located at $ u\rightarrow \infty $. Eq.(\ref{E2}) is precisely the starting point of our analysis. It is interesting to note that in the extremal limit, the near horizon structure corresponding to the above class of charged black branes turns out to be $ AdS_{2} \times R^{3} $ which therefore suggests that holographically the IR physics of these quantum critical points is described in terms of  a one dimensional conformal field theory ($ CFT_1 $) where only the time coordinate scales with $ z=\infty $, where $ z $ is the dynamic critical exponent \cite{Edalati:2012tc}. 

The physical picture behind our analysis is the following. The heavy charged quark at the boundary could be thought of as the end point of an open string stretched in between a \textit{probe} $ D $ brane\footnote{The probe $D$- brane essentially corresponds to the limit $ N_f \ll N $, where $ N_f $ corresponds to the number of hypermultiplates in the fundamental representation of the gauge group \cite{Fischler:2014tka}. Under this limiting condition one can therefore ignore the effect of backreaction on the background space time due to flavor branes. } (that is radially extended from the boundary at $ r\rightarrow \infty $ to a finite radial distance at $ r =r_B $) and the horizon at $ r=r_H $ \cite{Edalati:2012tc}. We then apply an external electric field $ \mathfrak{F}_{xt}=\mathfrak{F}(t)\sim e^{i \mathfrak{w}t}\mathfrak{F}(\mathfrak{w}) $ on the flavor branes which thereby exerts some electric force on the charged quark and thereby making it fluctuate. Such an external force could in principle be realized in terms of a boundary action of the form,
\begin{eqnarray}
S_{EM}=\int dt dr \delta(r-r_B) \mathfrak{F}(t)X(t,r)
\end{eqnarray} 
which thereby does not affect the bulk equation of motion. We then study the linear response of the quantum critical system to the fluctuations associated with the massive quark at the boundary. In order to study these fluctuations one essentially needs to go beyond the usual classical description of the string and consider fluctuations over the induced worldsheet geometry. These fluctuations could be thought of as propagating scalar modes over the string worldsheet. What we essentially compute is the low frequency behavior of the two point correlation between these scalar modes at the boundary. 

The dynamics of the string is described in terms of the Nambu-Goto action,
\begin{eqnarray}
S_{NG}= -\frac{1}{2 \pi \alpha'}\int d^{2}\sigma \sqrt{- det \gamma_{ab}}\label{E3}
\end{eqnarray}
where $ \gamma_{ab}=g_{\mu\nu}\partial_a X^{\mu}\partial_b X^{\nu} $ is the induced metric on the string worldsheet and $ X^{\mu} $s are the so called string embedding functions \cite{Tong:2012nf}-\cite{Edalati:2012tc}. Expanding the above action (\ref{E7}) upto quadratic order in the fluctuations we note that\footnote{In the present analysis we set our static gauge as, $ \tau =t $ and $ \sigma =u$. Moreover, as the fluctuations essentially decouple from each other therefore in our analysis we consider fluctuations along a particular direction which for the present case is along the $ x $ direction. Note that by doing so we effectively focus our attention to the three dimensional slice of the asymptotic AdS spacetime.},
\begin{eqnarray}
S_{NG}\approx S_{NG}^{(0)}-\frac{1}{4 \pi \alpha'}\int dt~ du \left[\frac{r_H^{3}\mathcal{N}u^{4}f(u)}{l^{4}}\left(\frac{\partial x}{\partial u} \right)^{2} -\frac{r_H}{2 \mathcal{N}f(u)}\left(\frac{\partial x}{\partial t} \right)^{2} \right]\label{E4}
\end{eqnarray}
where the first term on the r.h.s of (\ref{E4}) corresponds to the classical ground state of the static string\footnote{Holographically, for the present case, a static straight string hanging from $ r=r_B $ down to the black brane horizon at $ r=r_H $ corresponds to a static charged (quark like) particle at rest at the boundary.} with the energy density,
\begin{eqnarray}
E =\frac{\mathcal{N}(r_B - r_H)}{2 \pi \alpha'}. 
\end{eqnarray} 
Note that if we extend the string all the way to $ r \rightarrow \infty $ then the quark becomes infinitely massive and thereby loses all its dynamics at the boundary. Therefore in order to have a finite mass for the quark one needs to fix a cutoff $ r=r_B $ which is precisely the location of the flavor brane.

The equation of motion corresponding to these fluctuations turns out to be,
\begin{eqnarray}
\frac{\partial}{\partial u}\left(\frac{r_H^{3}\mathcal{N}u^{4}f(u)}{l^{4}}\left(\frac{\partial x}{\partial u} \right) \right)-\frac{r_H}{2 \mathcal{N}f(u)}\left(\frac{\partial^{2}x}{\partial t^{2}} \right)=0.\label{E6}  
\end{eqnarray}
Expressing these fluctuations in terms of the Fourier modes namely, $ x(u,t)\sim e^{-i\omega t} h_{\omega}(u)$ and substituting it back into (\ref{E6}) we obtain,
\begin{eqnarray}
\frac{\partial}{\partial u}\left(\frac{r_H^{3}\mathcal{N}u^{4}f(u)}{l^{4}}\left(\frac{\partial h_{\omega}}{\partial u} \right) \right)+\frac{\omega^{2}r_H}{2 \mathcal{N}f(u)}h_{\omega}=0.\label{E7}  
\end{eqnarray}

In order to obtain a systematic solution ($ h_\omega $) near the boundary of the space time, one needs to consider the above equation (\ref{E7}) separately in the following three regions.

%%%%%%%%%%%%%%%%%%%%%%%%%%%%%
\subsubsection{Near horizon region ($ u \sim 1 $)}
In order to solve the above equation in the near horizon limit we consider the following ansatz namely \footnote{We have replaced $ \omega\rightarrow \mathfrak{w}/l $ and $ \xi \rightarrow \tilde{\xi}/l $ where $ \mathfrak{w} $ and $ \tilde{\xi} $ are dimensionless entities.},
\begin{eqnarray}
u=1+\frac{\mathfrak{w}}{\tilde{\xi}}\label{E8}
\end{eqnarray}
such that in the limit $ u\rightarrow 1 $ both $ \mathfrak{w}\rightarrow 0 $ and $ \tilde{\xi}\rightarrow 0 $ in such a way so that the ratio $\mathfrak{w}/\tilde{\xi} \rightarrow 0 $ \cite{Edalati:2009bi}.
Using (\ref{E8}), the corresponding equation for these fluctuations turns out to be,
\begin{eqnarray}
\frac{\partial^{2}h_{\mathfrak{w}}}{\partial \tilde{\xi}^{2}} + \kappa^{2}h_{\mathfrak{w}} \approx 0\label{E9}
\end{eqnarray}
where, $ \kappa^{2} = \frac{l^{2}}{144r_H^{2}(1+\sqrt{1-4\lambda})} $. 

Finally, the most general solution corresponding to (\ref{E9}) could be read off as,
\begin{eqnarray}
h^{(in/out)}_{\mathfrak{w}} = e^{\pm i\kappa \tilde{\xi}}\approx 1 \pm \frac{i\kappa \mathfrak{w}}{(u-1)}.\label{E10}
\end{eqnarray}
Here the superscript (in)/(out) corresponds to ingoing and outgoing waves with respect to black hole horizon in the bulk space time. At this stage it is customary to mention that the notion of ingoing or outgoing modes are perfectly valid for the worldsheet scalar modes. This is due to the fact that the worldsheet metric in the absence of any fluctuation could be expressed in the form of a black hole metric whose horizon precisely coincides with that of the black hole horizon in the bulk. As a result, one can go a step further and define the so called tortoise coordinate in order to distinguish between the so called ingoing and outgoing modes with respect to these worldsheet fluctuations \cite{Fischler:2014tka}. 

%%%%%%%%%%%%%%%%%%%%%%%%%%%%%%
\subsubsection{Intermediate region $ (1<u<\infty) $}
In order to solve (\ref{E7}) in the intermediate region we consider the following power series expansion in the frequency $ \mathfrak{w} $ namely,
\begin{eqnarray}
h_{\mathfrak{w}}=h_{\mathfrak{w}}^{(0)}+\mathfrak{w}^{2}h_{\mathfrak{w}}^{(2)}+ \mathcal{O}(\mathfrak{w}^{3}).\label{E11}
\end{eqnarray} 

As a next step, we substitute (\ref{E11}) into (\ref{E7}) and consider the equation corresponding to $h_{\mathfrak{w}}^{(0)}$ which turns out to be,
\begin{eqnarray}
\frac{\partial}{\partial u}\left(\frac{r_H^{3}\mathcal{N}u^{4}f(u)}{l^{4}}\left(\frac{\partial h_{\mathfrak{w}}^{(0)}}{\partial u} \right) \right)=0.\label{E12}
\end{eqnarray}
%\begin{figure}[h]
%\centering
%\rotatebox{270}{
%\includegraphics[angle=0,width=10cm,keepaspectratio]{fig.pdf}
%\caption[]{\it Radial function ($ f(u) $) plot for $\lambda =0.05 $ with respect to the radial distance ($u$).}
%\label{figure 2a}
%\end{figure}

At this stage one should take a note on the fact that it is indeed a difficult task to solve the above equation (\ref{E12}) exactly. One therefore needs to make certain clever approximations. For certain choice of the coupling ($ \lambda $), it is in fact quite trivial to check that the radial function $ f(u) $ quickly saturates to its boundary value as one moves slightly away from the horizon. Therefore for our purpose it is indeed sufficient to check the solution within that range. Considering all these facts, the corresponding solution in the arbitrary region ($ u>1 $) turns out to be,
\begin{eqnarray}
h_{\mathfrak{w}}^{(0)}&=& \mathcal{C}_{2}-\frac{\mathcal{C}_1}{12(u-1)}
+\mathcal{C}_1\left( - \frac{5776}{(144 \lambda +211)^3 u}+\frac{392 u+38}{(144 \lambda +211)^2 u^2}-\frac{6 u (5 u+1)+1}{3 (144 \lambda +211) u^3}\right)+\mathcal{C}_1 \mathcal{F}(u)\nonumber\\
 \mathcal{F}(u) &=& \frac{1}{36} \log (1-u)+\frac{\left( \mathbb{Z}+\mathfrak{X}(\lambda)\tan ^{-1}\left(\frac{144 \lambda  (u-1)+147 u-173}{8 \sqrt{27 \lambda +17}}\right) \right) }{\sqrt{27 \lambda +17} (144 \lambda +211)^4 }\nonumber\\
 \mathbb{Z} &=& -4 \sqrt{27 \lambda +17}\left(\mathfrak{R}(\lambda)\log (u)+\mathfrak{Q}(\lambda)\log \left(147 u^2+144 \lambda  (u-1)^2-346 u+211\right) \right) \label{E13}
 \end{eqnarray}
where $ \mathfrak{X}(\lambda) $, $ \mathfrak{R}(\lambda) $ and $ \mathfrak{Q}(\lambda) $
are some complicated polynomials in $ \lambda $ whose detail we skip as they are not quite illuminating. Expanding the above solution (\ref{E13}) around $ u \sim 1 $ and comparing the $ \mathcal{O}(u-1)^{-1} $
as well as the $ \mathcal{O}(1) $ terms with solution (\ref{E10}) we note that,
\begin{eqnarray}
\mathcal{C}^{(in/out)}_2 =1- \mathcal{C}^{(in/out)}_1\mathbb{G} (\lambda),~~\mathcal{C}^{(in/out)}_1 =\mp 12 i \kappa \mathfrak{w}
\end{eqnarray}
where $ \mathbb{G} (\lambda) $ is a complicated polynomial of the following form, 
\begin{eqnarray}
\mathbb{G}(\lambda)&=&\frac{\mathfrak{A}(\lambda)}{288 \sqrt{27 \lambda +17} (144 \lambda +211)^4}\nonumber\\
\mathfrak{A}(\lambda)&=&(144 \lambda ~ \mathfrak{M}(\lambda) +354545440087)\tan ^{-1}\left(\frac{13}{4 \sqrt{27 \lambda +17}}\right)-4 \sqrt{27 \lambda +17}~(\mathfrak{N}_{1}(\lambda)+\mathfrak{N}_{2}(\lambda))
\nonumber\\
\mathfrak{M}(\lambda)&=&2592 \lambda  (16 \lambda  (72 \lambda  (432 \lambda +2587)+440819)+8165535)+11800193911\nonumber\\
\mathfrak{N}_{1}(\lambda)&=&24 (144 \lambda +211) (576 \lambda  (1332 \lambda +3581)+1392415)+288 \lambda  (72 \lambda +193)
\nonumber\\
\mathfrak{N}_{2}(\lambda)&=&(288 \lambda  (72 \lambda +193)+37837)(288 \lambda  (72 \lambda +589)+186709) \log (12).
\end{eqnarray}

On the other hand, expansion of the above solution (\ref{E13}) corresponding to the large values of $ u $ yields,
\begin{eqnarray}
h_{\mathfrak{w}}^{(0)}\approx \mathcal{C}_2  -\frac{\mathcal{C}_1 (432 \lambda  (144 \lambda  (48 \lambda +251)+59833)+13813219)}{12 (144 \lambda +211)^3 u}-\frac{96 \mathcal{C}_1 (3 \lambda +4)}{(144 \lambda +211)^2 u^2}.
\label{E16}
\end{eqnarray}

%%%%%%%%%%%%%%%%%%%%%%%%%%%%%%%
\subsubsection{Asymptotic region $ (u\rightarrow\infty) $}
Near the asymptotic region, the functional form of the metric $ f(u) $ turns out to be,
\begin{equation}
f_{\infty}=\frac{1}{2\lambda}\left[1-\sqrt{1-4\lambda} \right].
\end{equation}

As a result, the asymptotic form of the original equation (\ref{E7}) turns out to be,
\begin{eqnarray}
\frac{\partial}{\partial u}\left(u^{4}\left(\frac{\partial h_{\mathfrak{w}}}{\partial u} \right) \right)+\frac{\mathfrak{w}^{2}l^{2}}{2r_H^{2} \mathcal{N}^{2}f_{\infty}}h_{\mathfrak{w}}=0.\label{E18}
\end{eqnarray}
The solution corresponding to (\ref{E18}) could be formally expressed as,
\begin{eqnarray}
 h_{\mathfrak{w}}\approx \mathfrak{C}_{2}-\frac{i\sqrt{2} \mathfrak{C}_{1}}{\mathfrak{w} ^3 \left(\frac{l^2}{\mathcal{N}^2 f_{\infty } r_H^2}\right){}^{3/2}}-\frac{i\mathfrak{C}_{1}}{2 \sqrt{2} \mathfrak{w}  \sqrt{\frac{l^2}{\mathcal{N}^2 f_{\infty } r_H^2}}} \frac{1}{u^{2}}.  \label{E19} 
\end{eqnarray}
Finally, comparing (\ref{E19}) with (\ref{E16}) and retaining terms upto quadratic order in the frequency ($ \mathfrak{w} $) we note that,
\begin{eqnarray}
\mathfrak{C}^{(in/out)}_{1}&= & \mp \frac{2304\sqrt{2}\kappa \mathfrak{w}^{2} l (3\lambda +4)}{\mathcal{N}\sqrt{f_{\infty}}r_H(144\lambda +211)^{2}} \nonumber\\
\mathfrak{C}^{(in/out)}_{2}&=&\frac{i\sqrt{2} \mathfrak{C}^{(in/out)}_{1}}{\mathfrak{w} ^3 \left(\frac{l^2}{\mathcal{N}^2 f_{\infty } r_H^2}\right){}^{3/2}}+ \mathcal{C}^{(in/out)}_{2}
\end{eqnarray}

With the above solution (\ref{E19}) in hand, we next proceed to calculate the linear response of the system due to the presence of an external force $ \mathfrak{F} $. From the bulk point of view, the notion of force could be thought of as turning on a world volume $ U(1) $ gauge field for the probe brane. As the end point of the string is charged due to this $ U(1) $ field, therefore in order to add the desired force at the end point of the string, we add an additional term to the Nambu-Goto action (\ref{E3}) which is of the form,
\begin{eqnarray}
S_{EM}=\int_{\partial M} dt (A_t +A_{x}\dot{x} )
\end{eqnarray}
which is a pure boundary term and therefore does not affect the bulk equation of motion.

 The linear response of the system in the presence of an external source could be expressed as \cite{Edalati:2012tc}, 
\begin{eqnarray}
<x_{\mathfrak{w}}> = \mathfrak{G} (\mathfrak{w})\mathfrak{F} (\mathfrak{w})\label{E21}
\end{eqnarray}
where $\mathfrak{G} (\mathfrak{w})$ is the retarded Green's function and/or the admittance of the system. Knowing the retarded Green's function, our goal would be to compute the two point quantum correlation  for the boundary theory which is directly related to the imaginary part of the retarded Green's function as \cite{Edalati:2012tc},
\begin{eqnarray}
< X(\mathfrak{w})X(0)> = 2~ \mathcal{I}\mathit{m} ~\mathfrak{G} (\mathfrak{w}).
\end{eqnarray}

In order to compute the external force at the boundary we consider the ingoing wave boundary condition near the horizon of the black brane (\ref{E2}) namely,
\begin{eqnarray}
x_{\mathfrak{w}} (t,u)= e^{-i\mathfrak{w} t } \mathfrak{B}_{\mathfrak{w}} h^{(in)}_{\mathfrak{w}} 
\end{eqnarray}
where $\mathfrak{B}_{\mathfrak{w}}$ is the normalization constant that is fixed by the normalization condition \cite{Tong:2012nf},
\begin{eqnarray}
<x_{\mathfrak{w}}|x_{\mathfrak{w'}}> = \delta (\mathfrak{w}-\mathfrak{w'}).
\end{eqnarray}

 Using the \textit{ingoing} wave boundary condition, the external force exerted on the fluctuating string could be formally read off as \cite{Edalati:2012tc},
 \begin{eqnarray}
 \mathfrak{F} (\mathfrak{w})\approx \frac{1152 i \mathfrak{w} \mathcal{N}r_B r_H^{2} \kappa \mathfrak{B}_{\mathfrak{w}}(3\lambda +4)}{\pi \alpha' l^{3}(144\lambda +211)^{2}}.\label{E23}
\end{eqnarray}  
Using (\ref{E19}), (\ref{E21}) and (\ref{E23}) and considering the limits $ \mathfrak{w}/r_H \ll 1 $ and $ r_B/r_H \gg 1 $ the two point function finally turns out to be,
\begin{eqnarray}
< X(\mathfrak{w})X(0)> ~ &\sim &~ \frac{1}{\mathfrak{w}}\frac{(144\lambda +211)^{2}}{ \mathcal{N} \kappa(3\lambda +4)}. \label{E24}
\end{eqnarray}
Eq (\ref{E24}) essentially represents the full \textit{non perturbative} $ \lambda $ corrected expression for the two point correlation. In order to see the leading order effects in the GB coupling, we expand the above expression (\ref{E24}) perturbatively upto quadratic order in $ \lambda $ which yields,
\begin{eqnarray}
< X(\mathfrak{w})X(0)> ~ \sim ~ \frac{1}{\mathfrak{w}}\left( 1+\frac{519\lambda}{844}+\frac{3249\lambda^{2}}{712336}\right).\label{E27} 
\end{eqnarray}
Eq.(\ref{E27}) gives an exact analytic expression for the modified two point quantum correlation function in the presence of higher derivative (GB) corrections to the bulk gravitational action. Like in the case for the usual two derivative theory of gravity, the $ 1/\mathfrak{w} $ factor sitting in front of the above expression in (\ref{E27}) essentially confirms the fact that the IR physics of the dual QFT is essentially controlled by an one dimensional conformal field theory ($ CFT_1 $) which could be thought of as a quantum critical point with $ z=\infty $ \cite{Edalati:2012tc}. It is also noteworthy to mention that in the limit of the vanishingly coupling ($ \lambda \rightarrow 0 $), our result (\ref{E27}) smoothly matches to that with the earlier findings in \cite{Edalati:2012tc}. The subleading corrections appearing in (\ref{E27}) essentially correspond to $ 1/\lambda_{t} $ corrections on the gauge theory side.  Interestingly enough we note that the subleading corrections are all additive (at least upto quadratic order in the GB coupling ($ \lambda $)) which therefore suggests that the higher derivative corrections always enhance the fluctuations around the quantum critical point and thereby increases the possibility of the quantum phase transition.

%%%%%%%%%%%%%%%%%%%%%%%%%%%%%%%%%
\subsection{Lovelock gravity}
%%%%%%%%%%%%%%%%%%%%%%%%%
\subsubsection{Preliminaries}
For the sake of completeness, it is customary to explore the fate of quantum critical points under a general holographic set up where the dual gravitational counterpart consists of most generic higher derivative corrections beyond the usual GB term. It is a well known fact that the presence of higher derivative terms in the gravitational action in general introduce ghosts in the theory and thereby violates unitarity. However, it is Zwiebach \cite{Zwiebach:1985uq} and Zumino \cite{Zumino:1985dp} who first pointed out that one could avoid the unitarity problem if the so called higher derivative terms are expressed as the dimensional continuations of the Euler densities. These are precisely the theories which are known as Lovelock gravity \cite{deBoer:2009gx}. For a brief review on black holes in Lovelock gravity see \cite{Garraffo:2008hu}.

Lovelock gravity is the most general classical theory of gravity which yields field equations for the metric at most at the two derivative level \cite{deBoer:2009gx}. The most generic Lovelock action in $ d+1 $ dimensions could be expressed as \cite{deBoer:2009gx},
\begin{eqnarray}
S = \int d^{d+1}x \sqrt{-g}\sum_{p=0}^{p\le \left[ \frac{d}{2}\right] } \alpha_{p}\mathfrak{L}_p \label{E62}
\end{eqnarray}
where  $ \left[ \frac{d}{2}\right] $ stands for the integral part of $ d/2 $ and $ \alpha_{p} $ is the $ p $ th order Lovelock coefficient. Here $ \mathfrak{L}_p $ s are the so called Euler densities,
\begin{eqnarray}
\mathfrak{L}_p= 2^{-p}\delta^{a_{1}b_{1}..a_{p}b_{p}}_{c_{1}d_{1}..c_{p}d_{p}}~R^{c_{1}d_{1}}\ _{a_{1}b_{1}}..R^{c_{p}d_{p}}\ _{a_{p}b_{p}}
\end{eqnarray}
where, $ \delta^{a_{1}b_{1}..a_{p}b_{p}}_{c_{1}d_{1}..c_{p}d_{p}} $ is the totally antisymmetric product of Kronecker delta symbols in both set of indices. It is quite evident from the definition of the theory itself that the Euler density ($ \mathfrak{L}_p $) either vanishes or appears as a total derivative for $ p> \left[ \frac{d}{2}\right]$ and therefore does not affect the equation of motion. The Euler densities corresponding to $ p=0 $ and $ p=1 $ essentially stand for the cosmological constant and the usual Einstein-Hilbert term respectively. These are therefore the simplest example of Lovelock gravity in ($ 3+1 $) dimensions. The first non trivial contribution to the action (\ref{E62}) appears in ($ 4+1 $) dimensions for $ p=2 $, where one could add a specific combination of the Riemann tensor that results in the so called GB term.

We start with the following Lovelock action in $ D $ dimensions \cite{Ge:2009ac},
\begin{eqnarray}
S=\frac{1}{16 \pi G_D}\int d^{D}x \sqrt{-g}(-2 \Lambda +R+\alpha_{2}\mathcal{L}_{GB}+\alpha_{3}\mathcal{L}_{3}-4 \pi G_D F^{2})\nonumber\\
\mathcal{L}_{3}=2R^{abcd}R_{cdef}R^{ef}\ _{ab}+8R^{ab}\ _{cd}R^{ce}\ _{bf}R^{df}\ _{ae}+24 R^{abcd}R_{cdbe}R^{e}\ _{a}\nonumber\\
+3RR^{abcd}R_{cdab}+24 R^{abcd}R_{ca}R_{db}+16 R^{ab}R_{bc}R^{c}\ _{a}-12RR_{ab}^{2}+R^{3}
\end{eqnarray}
where, $ \mathcal{L}_{GB} $ is the standard GB Lagrangian (\ref{E1}) and $ F_{(2)} (=d A_{(1)})$ is the two form field strength. With the following choice of the parameters,
\begin{eqnarray}
\alpha_{2}=\frac{\alpha}{(D-3)(D-4)},~~\alpha_{3}=\frac{\alpha^{2}}{3(D-3)..(D-6)}
\end{eqnarray}
the corresponding charged black hole solution turns out to be \cite{Ge:2009ac},
\begin{eqnarray}
ds^{2}&=&- \mathcal{H}(u)\mathfrak{V}^{2} dt^{2}+\mathcal{H}^{-1}(u)du^{2}+u^{2}dx_{i}^{2}\nonumber\\
\mathcal{H}(u)&=&\frac{u^{2}}{\alpha}\left[ 1-\left\lbrace 1- 3\alpha\left( 1-\frac{\mathfrak{a}+1}{u^{D-1}}+\frac{\mathfrak{a}}{u^{2D-4}}\right) \right\rbrace^{1/3} \right] \nonumber\\
\mathfrak{V}^{2} &=& \frac{\alpha}{1-(1-3\alpha)^{1/3}},~~\mathfrak{a}=\mathfrak{q}^{2}\label{E66}
\end{eqnarray}
where we have set the AdS length scale ($ l $) equal to unity. The radial coordinate $ u $ is defined as earlier such that the horizon is located at $ u=1 $ and the boundary is located at $ u \rightarrow \infty $. From the structure of the above solutions (\ref{E66}), it is indeed quite evident that in order to preserve the asymptotic AdS boundary conditions the corresponding value of the higher derivative coupling must be bounded above namely, $ \alpha \le 1/3 $. We will have more discussions about this bound in the subsequent sections.

%%%%%%%%%%%%%%%%%%%%%%%%%%%%%%%%%%%%%%%%%%%%%%
\subsubsection{Extremal limit}
In the present calculation we would be interested in the so called extremal limit of the above black hole solution (\ref{E66}). The final object that we want to evaluate is the two point correlation function at $ T=0 $. In our analysis we would set $ D=7 $. The extremal limit is achieved by setting $ \mathfrak{a}=\mathfrak{q}^{2}=3/2 $ \cite{Ge:2009ac} which yields,
\begin{eqnarray}
\mathcal{H}(u)=\frac{u^{2}}{\alpha}\left[ 1-\left\lbrace 1- 3\alpha\left( 1-\frac{5}{2}~u^{-6}+\frac{3}{2}~u^{-10}\right) \right\rbrace^{1/3} \right].\label{E67}
\end{eqnarray}

Before we proceed further, let us first explore the near horizon structure of \textit{extremal} charged black brane solutions in third order Lovelock gravity as mentioned above in (\ref{E67}). To explore the near horizon structure we set,
\begin{eqnarray}
u=1+\varepsilon, ~|\varepsilon| \ll 1.
\end{eqnarray}
For our analysis, it is in fact enough to focus only on the $ (t-u) $ sector of the metric which turns out to be,
\begin{eqnarray}
ds^{2}\sim -\varepsilon^{2}dt^{2}+\frac{d\varepsilon^{2}}{\varepsilon^{2}}+..~..
\end{eqnarray}
Finally, by setting $ \gamma =1/\varepsilon $ we arrive at,
\begin{eqnarray}
ds^{2}= \frac{1}{\gamma^{2}}\left( -dt^{2}+d\gamma^{2}\right) +..~.. \equiv AdS_{2} \times \mathbb{R}^{5}.\label{e70}
\end{eqnarray}
From (\ref{e70}), it is quite evident that the boundary field theory in its deep IR flows towards a one dimensional $ CFT_{1} $ dual to $ AdS_{2} $. Like we noticed earlier, it is a quantum critical point with $ z= \infty $ which thereby suggests that in the low frequency limit, the imaginary part of the admittance should scale as the inverse of the frequency ($ \mathfrak{w} $) itself.

The energy corresponding to the static string turns out to be,
\begin{eqnarray}
E = \frac{\mathfrak{V}(r_B -r_H)}{2 \pi \alpha^{'}r_H}.
\end{eqnarray}

Considering small fluctuations, the linearized equation of motion about the average configuration of the string turns out to be,
\begin{eqnarray}
\frac{\partial}{\partial u}\left(\mathcal{H}(u)u^{2}\left( \frac{\partial x}{\partial u}\right)  \right) - \frac{u^{2}}{\mathcal{H}\mathfrak{V}^{2}}\left( \frac{\partial^{2}x}{\partial t^{2}}\right) =0.
\end{eqnarray}
Substituting fluctuations in terms of their Fourier modes namely, $ x(t,r)=e^{-i \mathfrak{w}t}h_{\mathfrak{w}}(r) $ we finally arrive,
\begin{equation}
\frac{\partial}{\partial u}\left(\mathcal{H}(u)u^{2}\left( \frac{\partial h_{\mathfrak{w}}}{\partial u}\right)  \right) +\frac{u^{2}\mathfrak{w}^{2}}{\mathcal{H}\mathfrak{V}^{2}}h_{\mathfrak{w}}=0.\label{E70}
\end{equation}

%%%%%%%%%%%%%%%%%%%%%%%%%%%%%%
\subsubsection{Near horizon region ($ u \sim 1 $)}
In order to solve (\ref{E70}) in the near horizon limit we set,
\begin{eqnarray}
u = 1+\frac{\mathfrak{w}}{\varsigma}.\label{E71}
\end{eqnarray}
Using (\ref{E71}), the near horizon structure of (\ref{E70}) turns out to be
\begin{eqnarray}
h''_{\mathfrak{w}}(\varsigma)+\left( \frac{1}{30 \mathfrak{V}}\right) ^{2}h_{\mathfrak{w}}(\varsigma)=0.
\end{eqnarray}
The corresponding \textit{ingoing} solution turns out to be,
\begin{eqnarray}
h_{\mathfrak{w}}(u)=1+ \frac{i \mathfrak{w}}{30 \mathfrak{V}(u-1)}+\mathcal{O}(\mathfrak{w}^{2}).\label{E76}
\end{eqnarray}
%%%%%%%%%%%%%%%%%%%%%%%%%%%%%
\subsubsection{Intermediate region $ (1<u<\infty) $}
Following the arguments as mentioned earlier in (\ref{E11}), the equation corresponding to zeroth order in the fluctuations turns out to be,
\begin{eqnarray}
\frac{\partial}{\partial u}\left(\mathcal{H}(u)u^{2}\left( \frac{\partial h^{(0)}_{\mathfrak{w}}}{\partial u}\right)  \right)=0.\label{E77}
\end{eqnarray}
The exact near horizon structure of the solution corresponding to (\ref{E77}) turns out to be,
\begin{eqnarray}
h^{(0)}_{\mathfrak{w}}\approx \frac{7}{27} \mathfrak{m}_{1} \text{Ei}\left(\frac{7 (u-1)}{3}\right)+\frac{\mathfrak{m}_{1} e^{\frac{7 (u-1)}{3}}}{9-9 u}+\mathfrak{m}_{2}\label{E78}
\end{eqnarray}
where $ \text{Ei} (u)$ is the so called \textit{exponential integral} function \cite{Dover}.

From the near horizon structure (\ref{E76}), it should be clear by now that the unknown constant $ \mathfrak{m}_{1} $ could in principle be a linear function of the frequency ($ \mathfrak{w} $). 
Comparing (\ref{E76}) and (\ref{E78}) for $ u\sim 1 $, we finally note that,
\begin{eqnarray}
\mathfrak{m}_{1}=-\frac{3 i \mathfrak{w}}{10 \mathfrak{V}}.
\end{eqnarray}
Finally, the asymptotic solution corresponding to (\ref{E77}) could be formally expressed as,
\begin{eqnarray}
h^{(0)}_{\mathfrak{w}}\approx \mathfrak{m}_{2}-\frac{\mathfrak{m}_{1}}{3 u^3}.\label{E80}
\end{eqnarray}
%%%%%%%%%%%%%%%%%%%%%%%%%%%%%%%
\subsubsection{Asymptotic region $ (u\rightarrow\infty) $}
In the asymptotic limit ($ u \rightarrow \infty $), the functional form of $ \mathcal{H}(u) $ turns out to be,
\begin{eqnarray}
\mathcal{H}(u)=\frac{u^{2}}{\alpha}\left[ 1-\left\lbrace 1- 3\alpha \right\rbrace^{1/3} \right]=\frac{u^{2}}{\mathfrak{V}^{2}}.\label{E81}
\end{eqnarray}
Using (\ref{E81}), the asymptotic form of (\ref{E70}) turns out to be,
\begin{eqnarray}
h''_{\mathfrak{w}}+\frac{4}{u}h'_{\mathfrak{w}}+\frac{\mathfrak{V}^{2}\mathfrak{w}^{2}}{u^{4}}h_{\mathfrak{w}}=0
\end{eqnarray}
which has a solution,
\begin{eqnarray}
h_{\mathfrak{w}}=\frac{\mathfrak{n}_{2} (-\mathfrak{V})^{3/2}}{2 \mathfrak{w}^3 \mathfrak{V}^{9/2}}+\frac{\mathfrak{n}_{2}}{4 u^2 \mathfrak{w} \sqrt{-\mathfrak{V}^2}}+\left(\mathfrak{n}_{1}-\frac{\mathfrak{n}_{2}}{6 u^3}\right)+\frac{\mathfrak{n}_{2} \mathfrak{w} \sqrt{-\mathfrak{V}^2}}{16 u^4}+\mathcal{O}(\mathfrak{w}^{2}/u^{2}).\label{E83}
\end{eqnarray}
Comparing (\ref{E80}) and (\ref{E83}) we finally note that, 
\begin{eqnarray}
\mathfrak{n}_{2}=2 \mathfrak{m}_{1}=-\frac{3 i \mathfrak{w}}{5 \mathfrak{V}}.
\end{eqnarray}
Using (\ref{E83}), and computing the external force exerted at the end point of the string the two point correlation finally turns out to be,
\begin{eqnarray}
<X(\mathfrak{w})X(0)>\sim \frac{\mathfrak{V}^{-1}}{\mathfrak{w}}=\frac{1}{\mathfrak{w}}\left( 1+\frac{\alpha }{2}+\frac{17 \alpha ^2}{24}\right) +\mathcal{O}(\alpha^{3}).\label{E85}
\end{eqnarray}
As we noticed in the previous section, the above result in (\ref{E85}) merely reflects the fact that the dual field theory in its deep IR flows towards $ CFT_1 $ with $ z=\infty $ which is perfectly consistent with our earlier observation in (\ref{e70}). Moreover, like in the GB case, we notice that the holographic Lovelock corrections to the quantum two point correlation function are all additive at least upto quadratic order in the coupling which therefore (like the GB case) enhance the quantum fluctuations around the critical point. Finally, it should be noted that unlike the usual GB case \cite{Cai:2009zn}, in principle there does not exist any upper bound on the Lovelock coefficient ($ \alpha $) due to the fact that the effective speed of graviton wave packet ($ c_{g} $) is always less than the local speed of light ($ c=1 $) at the boundary \cite{Ge:2009ac}.

%%%%%%%%%%%%%%%%%%%%%%%%%%%%%%%%%%%%%%%%%%%%%%%%%%%%%%%%%%%%%%%%%%%
\section{Thermal diffusion}
The physics of thermal dissipation is actually encoded in certain parameter known as the coefficient of thermal diffusion ($ \mathfrak{D} $) that describes the so called Brownian motion of a heavy quark in a hot viscous plasma \cite{deBoer:2008gu}. All these analysis typically assume that the dynamics of the heavy quark in a strongly coupled $ \mathcal{N}=4 $ SYM plasma is governed by the so called Langevin equation which yields,
\begin{eqnarray}
\mathfrak{D}=\frac{4}{\sqrt{\lambda_{t}}}\frac{1}{2 \pi T}.\label{D}
\end{eqnarray}
From the knowledge of the diffusion constant (\ref{D}) one could in fact infer how strong the quark is coupled to the plasma, for example, the smaller value of the diffusion constant essentially corresponds to the stronger coupling as well as the shorter mean free path. 

The purpose of this particular part of our analysis is to turn our attention towards the physics of thermal dissipation corresponding to some Lifshitz like fixed point at finite coupling. The corresponding dual gravitational description in the bulk essentially consists of Gauss Bonnet (GB) corrected charged Lifshitz black holes in ($ 4+1 $) dimensions \cite{Pang:2009pd}. Holographically this construction in the bulk implies that we are basically looking for some sort of corrections to the boundary correlation function due to the presence of finite coupling on the gauge theory side, in the so called non relativistic limit.

As pointed out in \cite{Pang:2009pd}, it is indeed a quite difficult to obtain an exact GB corrected Lifshitz black hole solution in five dimensions. Therefore the author tried to find out a perturbative solution in the GB coupling ($ \lambda $). The action that one generally starts with is of the following form \cite{Pang:2009pd},
\begin{eqnarray}
S=\frac{1}{16 \pi G_{5}}\int d^{5}x\sqrt{-g}\left[R-2\Lambda - \frac{1}{4}F^{2}-\frac{m^{2}}{2}A^{2}-\frac{1}{4}\mathcal{F}^{2}+\frac{\lambda l^{2}}{2}\mathcal{L}_{GB} \right]. \label{e30}
\end{eqnarray}
Note that here the original two form fields $ F_{(2)} $ are auxiliary by construction and they have the only role in deforming the asymptotic geometry from AdS to Lifshitz. On the other hand, the original Lifshitz black hole is charged under the second $ U(1) $ gauge field $ \mathcal{A}_{(1)} $. Another intriguing fact about these charged Lifshitz black hole solutions is that they do not admit any extremal limit.

The GB corrected charged Lifshitz black holes that stems out from (\ref{e30}) could be formally expressed as \cite{Pang:2009pd},
\begin{eqnarray}
ds^{2}&=&l^{2}\left[ - r^{2z}f(r)g(r)dt^{2}+\frac{dr^{2}}{r^{2}f(r)}+r^{2}dx_{i}^{2}\right],~~(i=1,2,3 )\nonumber\\
f(r)&=&f_{0}(1+\lambda f_{0}),~~f_{0}(r)= 1 -\frac{\mathcal{Q}^{2}}{18 r^{6}},~~ g(r)= \exp\left( \frac{5 \lambda \mathcal{Q}^{2}}{27 r^{6}}\right).\label{E28}
\end{eqnarray}

In order to proceed further, we define a new variable $ u = r/r_H $ in terms of which the above metric (\ref{E28}) turns out to be,
\begin{eqnarray}
ds^{2}&=&l^{2}\left[ - (u r_H)^{2z}f(u)g(u)dt^{2}+\frac{du^{2}}{u^{2}f(u)}+u^{2}r_H^{2}dx_{i}^{2}\right],~~(i=1,2,3 )\nonumber\\
f(u)&=&f_{0}(1+\lambda f_{0}),~~f_{0}(u)= 1 -\frac{\mathcal{Q}^{2}}{18 u^{6}r_H^{6}},~~ g(u)= \exp\left( \frac{5 \lambda \mathcal{Q}^{2}}{27 u^{6}r_H^{6}}\right).\label{E29}
\end{eqnarray}
Note that the dynamic critical exponent corresponding to the above solution (\ref{E29}) is $ z=z_{0}+2\lambda (z_{0}-1) $, where $ z_{0}(=6 )$, is the dynamic exponent corresponding to the zero of the GB coupling ($ \lambda $).
The temperature of the boundary theory is precisely given by that of the Hawking temperature in the bulk namely \cite{Pang:2009pd},
\begin{eqnarray}
T = \frac{z_0 r_H}{4 \pi}\exp\left( \frac{5 \lambda}{3}\right). 
\end{eqnarray}

The holographic picture is essentially the same as what we have seen in the previous section. The massive particle and/or the heavy quark in this hot Lifshitz 
bath is represented by means of the end point of an open string suspended from the probe $ D $ brane placed at $ r=r_B $. The goal of our present analysis is to explore the effects of $ 1/N $ corrections on the Brownian motion of the charged particle in this hot Lifshitz bath. The quantity that we are finally interested to compute is the thermal diffusion constant which could be expressed in terms of retarded Green's 
function as \cite{Tong:2012nf},
\begin{eqnarray}
\mathfrak{D}=-\lim_{\mathfrak{w}\rightarrow 0}i\mathfrak{w} T \mathfrak{G}(\mathfrak{w}).\label{E31}
\end{eqnarray}

In order to compute the above quantity in (\ref{E31}), we follow the same steps like we did in the previous section. As a first step of our analysis, we note down the Numbo-Goto action expanded upto quadratic order in the fluctuations,
\begin{eqnarray}
S_{NG}\approx S_{NG}^{(0)}-\frac{l^{2}r_H^{2+z}}{4 \pi \alpha'}\int dt~ du\left[u^{z+3}\sqrt{g(u)}f(u)\left(\frac{\partial x}{\partial u} \right)^{2} -\frac{r_H^{-2z}u^{1-z}}{\sqrt{g(u)}f(u)}\left(\frac{\partial x}{\partial t} \right)^{2} \right].\label{E32}
\end{eqnarray}
As usual the leading term on the r.h.s of (\ref{E32}) represents the energy of the non fluctuating classical string which for the present case turns out to be,
\begin{eqnarray}
E\approx \frac{l^2 \left(r_B^6-r_H^6\right)}{12 \pi  \alpha' }+\frac{5 \lambda  l^2 \left(\mathcal{Q}^2 \log \left(\frac{r_B}{r_H}\right)+3 r_B^6 \left(6 \log \left(\frac{r_B}{r_H}\right)+6 \log \left(r_H\right)-1\right)+r_H^6 \left(3-18 \log \left(r_H\right)\right)\right)}{108 \pi  \alpha' }.
\end{eqnarray}
As usual, in the limit $ r \rightarrow \infty $ we have an infinite contribution to the rest energy of the quark which thereby makes it infinitely heavy. Therefore in order to have a meaningful notion on the Brownian movement of the particle one needs to place the probe brane at a finite radial distance.

The linearised equation of motion that follows directly from (\ref{E32}) could be formally expressed as, 
\begin{eqnarray}
\frac{\partial}{\partial u}\left(u^{z+3}\sqrt{g(u)}f(u)\left(\frac{\partial x}{\partial u} \right) \right)- \frac{r_H^{-2z}u^{1-z}}{\sqrt{g(u)}f(u)}\left(\frac{\partial^{2} x}{\partial t^{2}}\right)=0.\label{E34}
\end{eqnarray}
Substituting $x(u,t)\sim e^{-i\mathfrak{w} t} h_{\mathfrak{w}}(u)$ into (\ref{E34}) we finally obtain,
\begin{eqnarray}
\frac{\partial}{\partial u}\left(u^{z+3}\sqrt{g(u)}f(u)\left(\frac{\partial  h_{\mathfrak{w}}}{\partial u} \right) \right)+ \frac{\mathfrak{w}^{2}r_H^{-2z}u^{1-z}}{\sqrt{g(u)}f(u)}h_{\mathfrak{w}}=0.\label{E35}
\end{eqnarray}
In order to solve fluctuations $ h_{\mathfrak{w}} $, we follow exactly the same steps like we did in the previous section namely, we solve (\ref{E35}) in three different regions and obtain the exact solution near the asymptotic region by matching appropriately with the solution in the interpolating region.
%%%%%%%%%%%%%%%%%%%%%%%%%%%%%%%%%%%%
\subsubsection{Near horizon region ($ u \sim 1 $)}
In order to solve (\ref{E35}) in the near horizon ($ u\sim 1 $) limit we consider the following ansatz,
\begin{eqnarray}
u=1+\zeta, ~~ |\zeta | \ll 1.\label{E36}
\end{eqnarray}
Before we proceed further, let us first express (\ref{E35}) schematically as,
\begin{eqnarray}
\partial_u \left( \mathfrak{K}(u)h'_{\mathfrak{w}}(u) \right)+\frac{\mathfrak{w}^{2}}{\mathfrak{K}^{2}(u)}h_{\mathfrak{w}}= 0\label{E37}
\end{eqnarray}
where, $ \mathfrak{K}(u) \sim \sqrt{g(u)}f(u) $ very close to the horizon. Let us now confine ourselves upto leading order in the frequency ($ \mathfrak{w} $) for which the equation (\ref{E37}) simply reduces to,  
\begin{eqnarray}
\partial_u \left( \mathfrak{K}(u)h'_{\mathfrak{w}}(u) \right)\approx 0\label{E38}.
\end{eqnarray}
In the limit $ u \rightarrow 1 $, the function $ \mathfrak{K}(u) $ could be expanded as,
\begin{eqnarray}
\mathfrak{K}(u) = (u-1)\mathfrak{K}'(1)+..~~..\label{E39}
\end{eqnarray} 
Using (\ref{E36}) and (\ref{E39}), the above equation (\ref{E38}) trivially reduces to,
\begin{eqnarray}
h''_{\mathfrak{w}}(\zeta)+\frac{h'_{\mathfrak{w}}(\zeta)}{\zeta} \approx 0.\label{E40}
\end{eqnarray}
The solution corresponding to (\ref{E40}) (which is also consistent with the ingoing wave boundary condition \cite{Tong:2012nf}) could be formally expressed as,
\begin{eqnarray}
h_{\mathfrak{w}}(u)= \mathfrak{a}(1-i \mathfrak{w} \log(u-1)).\label{E41}
\end{eqnarray}
%%%%%%%%%%%%%%%%%%%%%%%%%%%%%%%%%%%%%
\subsubsection{Intermediate region $ (1<u<\infty) $}
Following the same steps as we did earlier, we would like to solve (\ref{E35}) for the intermediate region. In order to do that, we consider the perturbative expansion of the mode $ h_{\mathfrak{w}} $ in the frequency ($ \mathfrak{w} $) like we did earlier in (\ref{E11}). We consider the equation corresponding to the zeroth order mode in $ \mathfrak{w} $ namely,
\begin{eqnarray}
\frac{\partial}{\partial u}\left(u^{z+3}\sqrt{g(u)}f(u)\left(\frac{\partial  h^{(0)}_{\mathfrak{w}}}{\partial u} \right) \right)\approx 0.
\end{eqnarray}
The corresponding solution turns out to be,
\begin{eqnarray}
h^{(0)}_{\mathfrak{w}} (u)= \mathfrak{b}_{1}+\mathfrak{b}_{2} \int \frac{du}{u^{z+3}\sqrt{g(u)}f(u)}\label{E43}
\end{eqnarray}
where $ \mathfrak{b}_{1} $ and $ \mathfrak{b}_{2} $ are two constants of integration
that is to be fixed. The exact solution of (\ref{E43}) turns out to be rather difficult. Therefore instead of doing this integral exactly, we would like to estimate its value both near the horizon as well as at the asymptotic infinity. 

Let us first evaluate the integral in (\ref{E43}) in the near horizon approximation.  The near horizon expansion of $ \sqrt{g(u)}f(u) $ turns out to be,
\begin{eqnarray}
\sqrt{g(u)}f(u) \approx (u-1) \sqrt{g(1)}f'(1)+..~~..\label{E44}
\end{eqnarray}
Substituting (\ref{E44}) into (\ref{E43}) and performing the integral we finally obtain,
\begin{eqnarray}
h^{(0)}_{\mathfrak{w}} (u) \approx \mathfrak{b}_{1} + \frac{\mathfrak{b}_{2}}{\sqrt{g(1)}f'(1)} \log (u-1).\label{E45}
\end{eqnarray}
Comparing (\ref{E41}) and (\ref{E45}), we note,
\begin{eqnarray}
\mathfrak{b}_{1}= \mathfrak{a},~~~\mathfrak{b}_{2}=- i \mathfrak{w} \mathfrak{a}\sqrt{g(1)}f'(1).
\end{eqnarray}

We would now like to explore the solution (\ref{E43}) near the UV scale of the theory. Before we actually compute the integral in (\ref{E43}), it is important to note down the asymptotic behavior of the functions $ g(u) $ and $ f(u) $,
\begin{eqnarray}
g(u)&\approx &1+\frac{5 \lambda \mathcal{Q}^{2}}{27 u^{6}r_H^{6}}\nonumber\\
f_{0}(u)& = & 1-\frac{\mathcal{Q}^{2}}{18 u^{6}r_H^{6}}.\label{E47}
\end{eqnarray}
Using (\ref{E47}), the above solution (\ref{E43}) turns out to be,
\begin{eqnarray}
h^{(0)}_{\mathfrak{w}} (u) &\approx & \mathfrak{b}_{1}+\mathfrak{b}_{2}\left[ \frac{(\lambda  (5 \lambda -1)-3) \mathcal{Q}^2 u^{-z-8}}{54 (\lambda +1)^2 (z+8) r_H^6}-\frac{u^{-z-2}}{(\lambda +1) (z+2)}\right] .\label{E48}
\end{eqnarray}
Substituting the value corresponding to the dynamic exponent ($ z $) and expanding upto leading order in the GB coupling ($ \lambda $) we finally note that,
\begin{eqnarray}
h^{(0)}_{\mathfrak{w}} (u) &\approx & \mathfrak{b}_{1}+\mathfrak{b}_{2}\left[ \left(-\frac{\mathcal{Q}^2}{252 u^{14} r_H^6}-\frac{1}{8 u^8}\right)+\lambda  \left(\frac{5 \mathcal{Q}^2 (21 \log (u)+5)}{2646 u^{14} r_H^6}+\frac{40 \log (u)+9}{32 u^8}\right)\right] .\nonumber\\
\label{E49}
\end{eqnarray}

%%%%%%%%%%%%%%%%%%%%%%%%%%%%%%%%%%%
\subsubsection{Asymptotic region $ (u\rightarrow\infty) $}
We would now like to evaluate (\ref{E35}) directly at the asymptotic infinity. The asymptotic form of (\ref{E35}) turns out to be,
\begin{eqnarray}
h''_{\mathfrak{w}}+ \mathfrak{g}(u) h'_{\mathfrak{w}}+\mathfrak{w}^{2}\mathfrak{z}(u)h_{\mathfrak{w}}=0\label{E50}
\end{eqnarray}
where, the coefficients could be formally expressed as,
\begin{eqnarray}
\mathfrak{g}(u)&\approx &\frac{\left(-5 \lambda ^2+\lambda +3\right) \mathcal{Q}^2}{9 (\lambda +1) u^7 r_H^6}+\frac{z+3}{u}\nonumber\\
\mathfrak{z}(u)&\approx &\frac{\left(-5 \lambda ^2+\lambda +3\right) \mathcal{Q}^2 u^{-2 z-8} r_H^{-2 (z+3)}}{27 (\lambda +1)^3}+\frac{u^{-2 z-2} r_H^{-2 z}}{(\lambda +1)^2}\nonumber\\
&=& \frac{9 u^6 r_H^6+\mathcal{Q}^2}{9 u^{20} r_H^{18}}+\frac{\lambda}{u^{20}}(\#)+\mathcal{O}(\lambda^{2}).\label{E51}
\end{eqnarray}
It should be quite clear by now that in the large $ u $ limit\footnote{By large $ u $ limit we always mean $ r_B/r_H \gg 1 $.} of (\ref{E50}) the term associated with the frequency ( $ \mathfrak{w}\rightarrow 0 $) becomes insignificant if we neglect all the terms beyond $ \mathcal{O}(1/u^{8}) $. As a result, the solution corresponding to (\ref{E50}) turns out to be,
\begin{eqnarray}
h_{\mathfrak{w}} \approx  \mathfrak{d}_1 \left[ 1- 2^{\frac{z-4}{6}} 3^{z/2}\left(\frac{(\lambda +1) r_H^6}{(\lambda  (5 \lambda -1)-3) \mathcal{Q}^2}\right){}^{\frac{z+2}{6}} \Gamma \left(\frac{z+2}{6},\frac{\mathcal{Q}^2 (\lambda  (5 \lambda -1)-3)}{54 (\lambda +1) r_H^6}\right)\right] \nonumber\\
-\mathfrak{d}_{2}u^{-z-2} \left(\frac{(\lambda +1)  r_H^6}{(\lambda  (5 \lambda -1)-3) \mathcal{Q}^2}\right){}^{z/6}\frac{ \sqrt[3]{\frac{(\lambda +1) r_H^6}{(\lambda  (5 \lambda -1)-3) \mathcal{Q}^2}} \left(\frac{(\lambda  (5 \lambda -1)-3) \mathcal{Q}^2}{(\lambda +1) r_H^6}\right){}^{\frac{z+2}{6}}}{(z+2)}
\label{E52}
\end{eqnarray}
which could be schematically written as,
\begin{eqnarray}
h_{\mathfrak{w}}=\mathfrak{d}_1 \mathbb{M}(\lambda ,\mathcal{Q})-\mathfrak{d}_{2}\frac{u^{-z-2}}{(z+2)}\mathbb{N}(\lambda, \mathcal{Q}).\label{E53}
\end{eqnarray}

Comparing (\ref{E48}) and (\ref{E53}) we finally note that,
\begin{eqnarray}
\mathfrak{d}_1=\mathfrak{a}\mathbb{M}^{-1}(\lambda ,\mathcal{Q}),~~\mathfrak{d}_{2}=\frac{-i \mathfrak{w}\mathfrak{a}\sqrt{g(1)}f'(1)}{(\lambda +1)}\mathbb{N}^{-1}(\lambda, \mathcal{Q}).
\end{eqnarray}

Finally, the force acting at the end point of the string turns out to be,
\begin{eqnarray}
\mathfrak{F}(\mathfrak{w})|_{u \rightarrow \infty} \approx \frac{-i l^{2}r_H^{z+2}\mathfrak{w}\mathfrak{a}\sqrt{g(1)}f'(1)}{2 \pi \alpha' }.\label{E55}
\end{eqnarray}

Using (\ref{E55}), the imaginary part of the two point correlation turns out to be,
\begin{eqnarray}
Im \mathfrak{G}(\mathfrak{w}) = \frac{2 \pi \alpha'}{{\mathfrak{w} l^{2}r_H^{2+z}\sqrt{g(1)}f'(1)}}
\end{eqnarray}
which finally yields the coefficient of thermal diffusion (\ref{E31}) as,
\begin{eqnarray}
\mathfrak{D} &=& \frac{2 \pi \alpha' z_0^{2+z} }{(4\pi)^{2+z}l^{2}}\frac{e^{\frac{5 (z+2)\lambda}{3}}}{\sqrt{g(1)}f'(1)}T^{-1-z}\nonumber\\
& \sim & \left[ 1+ \frac{  \lambda \mathcal{Q}^{2}}{3 r_H^{6}} \left(  \frac{34 r_H^6}{\mathcal{Q}^2}+\frac{1}{18}\right) \right] T^{5-z} + \mathcal{O}(\lambda^{2}).\label{E60}
\end{eqnarray}
Two points are to be noted at this stage. Firstly, the GB corrections upto leading order is additive and Secondly, for $ z>5 $ the rate of diffusion decreases with temperature. On the other hand, for $ z<5 $ we have the usual scenario of increasing diffusion with temperature which clearly indicates a crossover at $ z=5 $. This feature is indeed quite different from the earlier observations on the \textit{uncharged} Lifshitz back ground \cite{Tong:2012nf} where the crossover was observed at $ z=2 $. Finally, in the limit $ \lambda \rightarrow 0 $ one recovers the corresponding result for the ordinary charged Lifshitz black holes in ($ 4+1 $) dimensions. From (\ref{E60}), it is in fact quite instructive to figure out some of the intriguing features about the dynamics of the Brownian particle in the hot Lifshitz bath. If we assume that the mean free path goes with the temperature as, $ L_{mfp}\sim T^{2-\frac{z}{2}} $ and the relaxation time has a temperature dependence $ \tau \sim 1/T  $, then the distance traveled by the Brownian particle in time $ t $ could be formally expressed as,
\begin{eqnarray}
\Delta X^{2}\sim \frac{L_{mfp}^{2}}{\tau}t = \mathfrak{D} t.
\end{eqnarray}
Therefore, one possible interpretation of the above result (\ref{E60}) comes from the fact that the mean free path of the Brownian particle scales differently in the presence of a non zero chemical potential and which shows up as a non trivial temperature scaling in the expression for the diffusion constant. 

%%%%%%%%%%%%%%%%%%%%%%%%%%%%%%%%%%%%%%%%%%%%%%%%%%%%%%%
\section{Summary and final remarks}
In the present paper, based on the methods developed in \cite{Edalati:2012tc}, we study the low frequency behavior of two point correlation function for a class of one dimensional CFTs corresponding to $ z=\infty $ fixed whose dual gravitational counterpart consists of extremal black holes in the usual higher derivative theories of gravity. The two point function thus evaluated at the end point of the open string has been found to receive some non trivial higher derivative corrections appearing from the gravity sector in the bulk. These corrections suggest that quantum fluctuations are enhanced due to the presence of higher derivative corrections on the gravity side which thereby increases the possibility of quantum transition around the quantum critical point. On the field theory side, such higher derivative corrections would correspond to an expansion in the coupling $ 1/\lambda_{t} $. It is also noteworthy to mention that our result smoothly matches to that with the earlier findings \cite{Edalati:2012tc} in the limit of the vanishing coupling.

Finally, for the sake of completeness and clarity, we compute the coefficient of thermal dissipation corresponding to Lifshitz like fixed points where the dual gravitational counterpart in the bulk consists of Gauss Bonnet corrected charged Lifshitz black hole solutions in ($ 4+1 $) dimensions. In our analysis we observe an important crossover corresponding to $ z=5 $ fixed point. We identify our result as a natural consequence of the non trivial scaling of the mean free path with temperature.

%%%%%%%%%%%%%%%%%%%%%%%%%%%%%%%%%%%%%%%%%%%%%%%%%%%%%%%%%%%%%%%%%%%%%%
{\bf {Acknowledgements :}}
 The author would like to acknowledge the financial support from CHEP, Indian Institute of Science, Bangalore.\\

%%%%%%%%%%%%%%%%%%%%%%%%%%%%%%%%%%%%%%%%%%%%%%%%%%%%%%%%%%%%%%%%%%%%%%%%%%%%%%%%%%%

\end{document}